\documentclass[conference]{IEEEtran}
\IEEEoverridecommandlockouts
\usepackage{amsmath,amssymb,amsfonts}
\usepackage{algorithmic}
\usepackage{graphicx}
\usepackage{textcomp}
\usepackage{xcolor}
\def\BibTeX{{\rm B\kern-.05em{\sc i\kern-.025em b}\kern-.08em
    T\kern-.1667em\lower.7ex\hbox{E}\kern-.125emX}}

\DeclareUnicodeCharacter{0308}{HERE!HERE!}

\begin{document}

\title{Automated Quantification of White Blood Cells in Light Microscopic Images of Injured Skeletal Muscle\\
}

\author{\IEEEauthorblockN{Yang Jiao$^{1}$, Hananeh Derakhshan$^{2}$, Barbara St. Pierre Schneider$^{2}$, Emma Regentova$^{1}$, Mei Yang$^{1}$}
\IEEEauthorblockA{\textit{1. Department of Electrical and Computer Engineering} \\
\textit{2. School of Nursing} \\
\textit{University of Nevada, Las Vegas}\\
Nevada, USA\\
jiaoy1@unlv.nevada.edu}

}

\maketitle

\begin{abstract}
White blood cells (WBCs) are the most diverse cell types observed in the healing process of injured skeletal muscles. In the course of healing, WBCs exhibit dynamic cellular response and undergo multiple protein expression changes. The progress of healing can be analyzed by quantifying the number of WBCs or the amount of specific proteins in light microscopic images obtained at different time points after injury. In this paper, we propose an automated quantifying and analysis framework to analyze WBCs using light microscopic images of uninjured and injured muscles. The proposed framework is based on the Localized Iterative Otsu’s threshold method with muscle edge detection and region of interest extraction. Compared with the threshold methods used in ImageJ, the LI Otsu’s threshold method has high resistance to no-object area and achieves better accuracy. The CD68-positive cell results are presented for demonstrating the effectiveness of the proposed work.
\end{abstract}


\section{Introduction}
Muscle injury commonly occurs during military training. Initial Military Training (IMT)-related musculoskeletal injuries seriously affect U.S. military recruits \cite{molloy2012physical}. Until 2012, during Basic Combat Training (BAT), approximately 25\% of male and 50\% of female recruits suffered from one or even more injuries \cite{bullockprevention}. These injuries are responsible for more than 80\% disability-related medical discharge among first-year recruits \cite{niebuhr2010accession,dekonig2006recruit}. The estimated cost per discharged recruit was \$57,500 in 2005 \cite{niebuhr2008assessment}. Therefore, understanding mechanisms promoting muscle recovery are crucial.

The injured muscle undergoes inflammatory response followed by muscle regeneration. As part of the inflammatory response to this injury, monocytes and macrophages initiate repair which includes the activation of skeletal muscle regeneration [6]. During this process, WBCs travel from the bone marrow to the injured muscle via the blood vessels. These WBCs exhibit a dynamic response by expressing a variety of proteins (e.g., CD68) at different post-injury time points \cite{dobek2013mouse}. The characteristics of WBCs can be analyzed by processing light microscopic images of uninjured muscle and muscle obtained at different time points after injury (4 to 192 hours). The basic analysis consists of cell recognition and counting.

Previous studies of muscle regeneration usually employ manual approach or basic intensity process to detect and count WBCs. However, manual analysis or manually cell counting has limitations. First, the success of the manual cell counting depends greatly on the image resolution \cite{schneider2010relation}. At higher resolutions, greater accuracy can be attained; however, this approach is laborious and thus costly. Under high resolution settings, multiple images have to be acquired to cover the whole injured area. Second, manual procedures are also error- prone due to the natural human fatigue after hours of analysis. Evidently, cell characteristics such as cell or cluster size are difficult to obtain manually. Third, it may take a long time to process images. Analysis of sophisticated microscopic muscle images needs rigor, which may take several rounds to check the results.

On the contrary, the computer vision (CV) techniques are promising for analyzing microscopic images, and therefore, are an alternative to the manual methods for biological research. Actually, in \cite{lejeune2008quantification}, M. Lejeune has achieved semi-automated quantification of diverse subcellular immunohistochemically markers with program Image-Pro Plus. However, it requires researcher to input complex commands at each step.

One of the central problems of computer vision is the segmentation of images, which is performed for obtaining regions of image related to different objects. Most existing work mainly focuses on high magnification image processing and WBC nucleus detection. For images with clustered WBCs and complicated background (such as the muscle edge, connective tissues and background stains), a new segmentation method that can adaptively and robustly process and analyze these images is needed.

Aiming to processing stained microscopic muscle images, we have developed a quantifying and analysis framework for automated WBC counting in images of both uninjured and injured muscles. The framework is based on the Localized Iterative Otsu’s threshold method with muscle edge detection and ROI extraction. The CD68-positive cell results are presented for demonstrating the effectiveness of the proposed work.

The rest of the paper is organized as follows. Section II reviews existing work on threshold based image segmentation methods. Section III introduces the quantifying and analysis framework. Section IV presents the Localized Iterative Otsu’s
method. Section V describes the muscle edge detection algorithm, region of interest (ROI) selection and its result. Section VI presents the analysis results of CD68-positive cells using the proposed framework. Section VII concludes the paper.

\section{Related Work}

In \cite{prinyakupt2015segmentation}, the processing scheme presented by the authors employs morphological and histogram processing. Specifically, adjacent objects can be separated by edge detection processing. Among existing threshold methods, one of the most efficient threshold methods is the Otsu’s method \cite{otsu1975threshold}, which automatically performs clustering-based image thresholding. Khobragade et al. demonstrate their method to detect leukemia WBCs. In their paper, histogram processing is employed with traditional Otsu’s threshold \cite{7489422}. Based on the gradient vector flow, Sadeghian et al. use active contour models (snakes) which are good to describe the contour of object and segment \cite{sadeghian2009framework}. In \cite{talebi2016automatic,gautam2022deep}, HoG and LBP descriptors are applied to train a classifier for detecting WBCs, which achieves 95.56\% accuracy.

An improved Otsu’s algorithm \cite{khan2016iterative, jiao2022fine} that iteratively evaluates the object size and changes threshold is developed by Khan et al. Their method resolves the uneven exposure problems. Ma et al. improve the Otsu’s method by drawing a clear boundary between classes of gray values \cite{sun2016improved, jiao2023digitally}. The work by Sa et al. combines Otsu’s algorithm with an edge detector to produce a more accurate segmentation \cite{sa2016improved,jiao2023learning,jiao2019deepquantify}. The 2D Otsu’s method is proposed in several papers as an effort to improve robustness of Otsu’s method in the presence of noises \cite{pan2016image,sha2015gray,teng2017feasibility,jiao2019multi}. Besides, applying 2D Otsu’s method iteratively can overcome noise and background distortion \cite{devi2015iterative,jiao2018automated,jiao2019self}. 2D Otsu’s method has been proved working well for bimodal Histograms.

The aforementioned work shows the validity of Otsu’s method for image segmentation. However, they mainly focus on high magnification image processing and WBC nucleus
detection. Our microscopic images are at 100x magnification, in which cell nuclei are not visible. In addition, complicated background including the muscle edge, connective tissues, and clustered cells make it hard to segment and analyze WBCs. We have used ImageJ, an open source Java image processing software \cite{schneider2012nih,chen2023improving}, to process these images. However, the threshold methods in ImageJ show poor performance for uninjured images. In our experiments, three threshold methods from ImageJ are selected to compare with the proposed LI Otsu’s threshold method.

\begin{figure}[h] 
    \centering
    \includegraphics[width=0.5\textwidth]{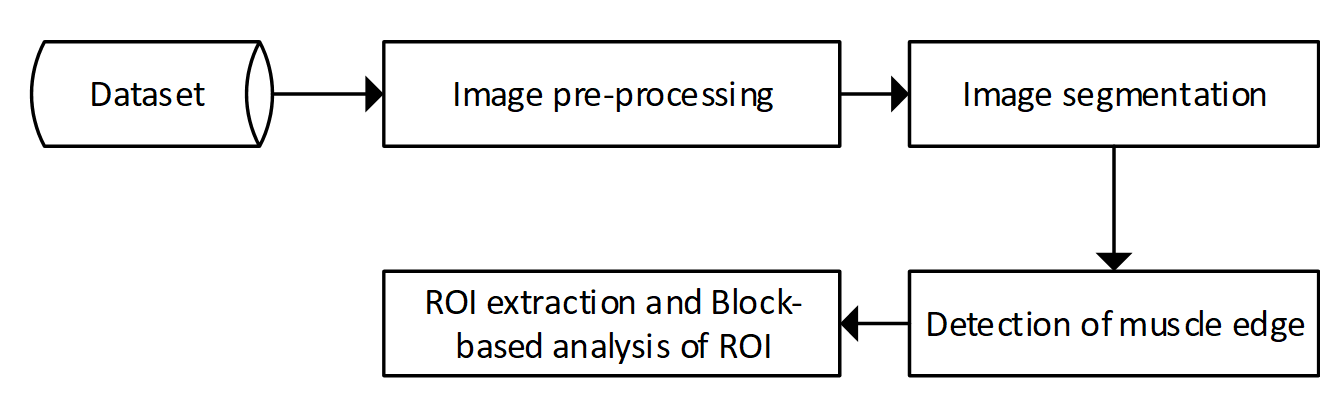} 
    \caption{Overview of the proposed framework}
    \label{fig:1}
\end{figure}

\begin{figure}[h] 
    \centering
    \includegraphics[width=0.5\textwidth]{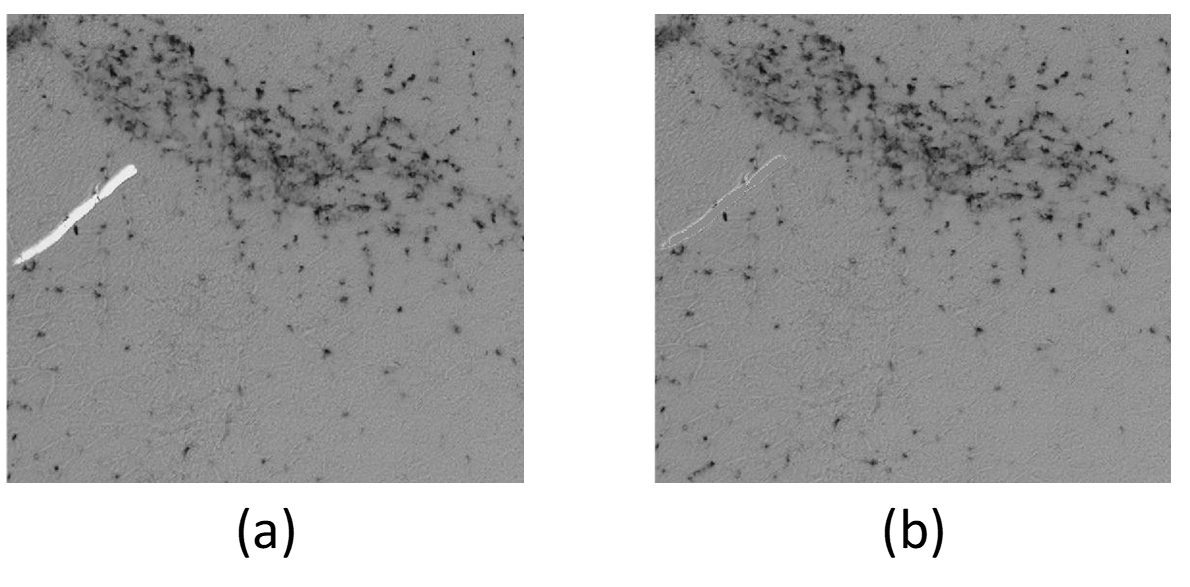} 
    \caption{(a) “White error” and (b) correctness}
    \label{fig:2}
\end{figure}

\section{AUTOMATED QUANTIFYING AND ANALYSIS FRAMEWORK}
The focus of this work is on segmentation of WBCs and their differentiation from other constituents of muscle images, such as muscle fibers, blood vessels, nerve bundles, and connective tissue. The input microscopic images are of resolution 1600x1200 pixels at 100x magnification. On each slide stained with antibody, WBCs are darker than other constituents; muscle fibers are clearly visible; WBCs can appear as discrete cells or clustered cells. A challenging issue for cell segmentation and quantification is that the cells can be concentrated at the muscle edge. Because the muscle edge often appears darker than other parts, it can be detected and excluded from consideration.

Fig. 1 shows the flowchart of the developed framework. It consists of four major steps: pre-processing, image segmentation using localized iterative Otsu’s threshold, muscle edge detection, and ROI selection and analysis.

\subsection{Pre-processing}
Pre-processing is applied to prepare images for further processing. An input image of 1600x1200 pixels is converted into the grayscale. Then, a global mean intensity value is calculated by Eq. (1).

\begin{equation}
\mu = (\Sigma_j^J\Sigma_k^Ki_{jk})/JK \label{eq1}
\end{equation}

Where $\mu$ is the mean pixel intensity, $J$ and $K$ represent the width and length of image respectively, $i$ is the intensity value of pixel at $(j, k)$.

Image artefacts such as air bubbles are corrected by filling in the pixel values with the calculated mean $\mu$. The input image is divided into blocks of 400x400 pixels and the local mean intensity value and the standard deviation $\sigma$ are found by Eq.
(1) and Eq. (2).

\begin{equation}
\sigma = \sqrt{(\Sigma_j^J\Sigma_k^K(i_{jk}-\mu_{local}))} \label{eq2}
\end{equation}

\subsection{Localized iterative Otsu’s threshold}
The Localized Iterative Otsu’s (LI Otsu’s) threshold is proposed to segment WBCs. Intensity variation within and across the images as well as contrast variation are common problems that may be encountered for the given type of imagery especially low resolution ones. Since the muscle fibers and the muscle edge can bias segmentation results, a local threshold is calculated. The implementation details of LI Otsu’s threshold method will be given in Section III.

\begin{figure}[h] 
    \centering
    \includegraphics[width=0.45\textwidth]{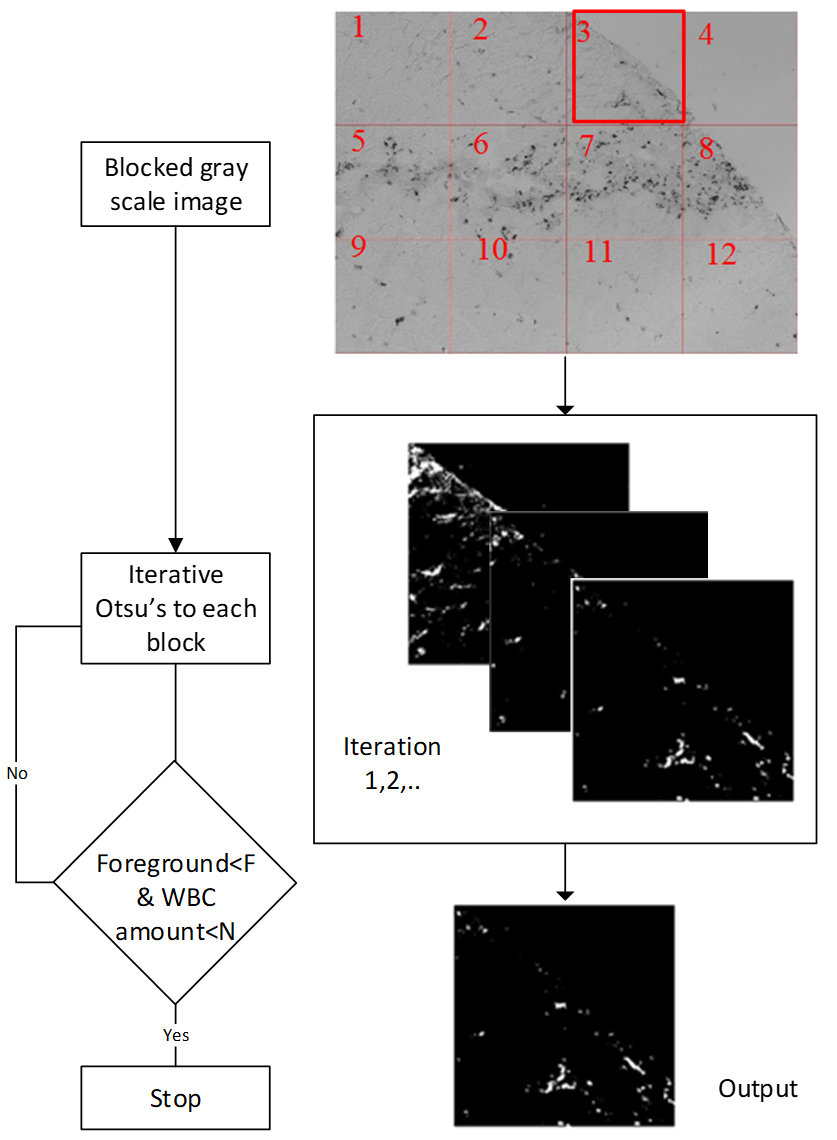} 
    \caption{Localized Iterative Otsu’s Threshold}
    \label{fig:3}
\end{figure}

\begin{figure}[h] 
    \centering
    \includegraphics[width=0.45\textwidth]{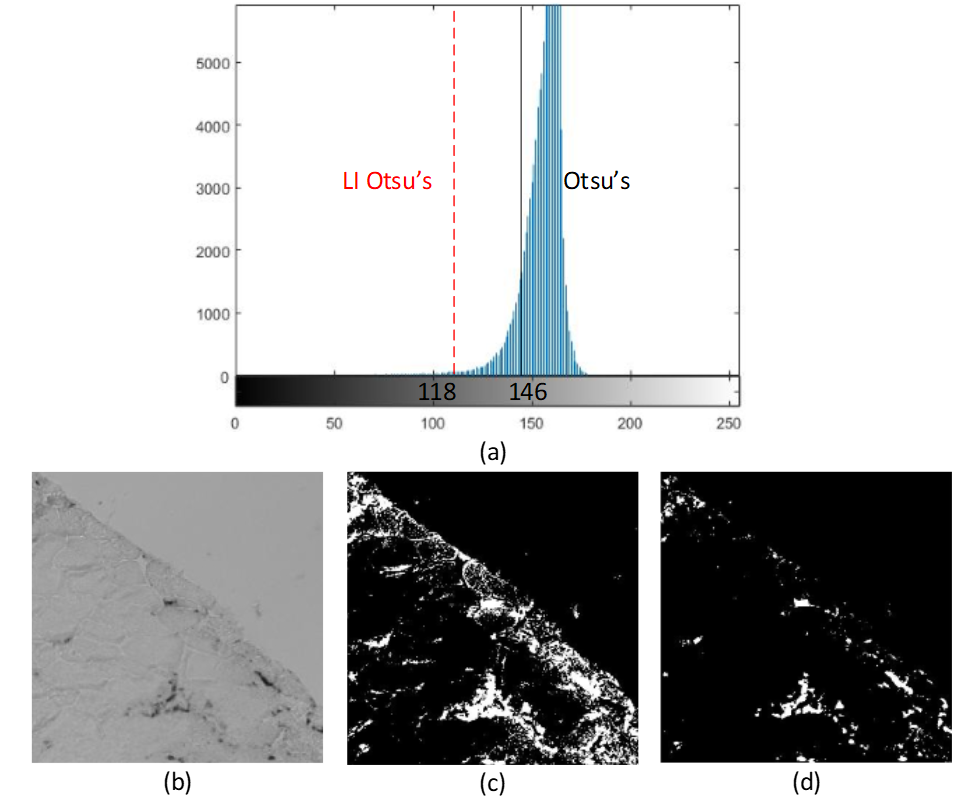} 
    \caption{(a) Histogram, (b) Gray scale image, (c) Otsu’s results, (d) LI Otsu’s results}
    \label{fig:4}
\end{figure}

\subsection{Muscle edge detection}

As discussed before, the muscle edge can interfere with WBC analysis because the edge may falsely acquire a dark color. Muscle edge detection is an important part of the proposed framework. In this step, the muscle edge is detected, and WBCs in close proximity to the muscle edge are excluded from subsequent analyses. Section IV describes the edge detection method.

\subsection{ROI selection and block-based analysis}

Via ROI selection, areas that are not related to muscle injury are excluded. Typically, ROIs shall not include the muscle edge and the surrounding tissues. To resolve the variation caused by uneven exposure, the ROI is divided into blocks of 400x400 pixels for block-based analysis. Given that an average size of cells is 500 pixels, a discrete cell or a cluster of cells can be differentiated. The following quantities are estimated:
\begin{itemize}
\item 1) Number of discrete WBCs
\item 2) Number of WBC clusters
\item 3) Average size of a discrete WBC
\item 4) Total area covered by WBC clusters
\item 5) Number of discrete cells in WBC clusters
\begin{equation}
N_{cic}=\frac{Area_{cluster}}{E[Size_{WBC}]} \label{eq3}
\end{equation}
\item 6) Total number of WBCs
\item 7) Number of WBCs per block
\item 8) Frequency of occurrence of a block with certain
number of WBCs (histogram).
\end{itemize}

\section{LOCALIZED ITERATIVE OTSU’S THRESHOLD}
For each 400x400 block, the local threshold is obtained by
applying the LI Otsu’s threshold method, as shown in Fig. 2.
Two parameters are pre-set: the expected foreground ratio $F$
and the expected object maximum count $N$. The value of $F$ and
$N$ are obtained by experiments. The method proceeds based on
Eq. (4) and Eq. (5).

\begin{equation}
\sigma_\omega^2 = \omega_0(t)\sigma_0^2(t) + \omega_1(t)\sigma_1^2(t), \label{eq4}
\end{equation}
where $t$ is a threshold, $\omega$ is the probability of class separation
by $t$, $\sigma^2$ is the class variance, $t$ is decided when $\sigma^2$ is
maximized [12]. If the calculated foreground ratio and the
maximum object area do not satisfy the pre-set $F$ and $N$
respectively, one more iteration is needed.

\begin{equation}
t_i = S*t_{i-1}, \label{eq5}
\end{equation}

where $i$ is an iteration number, $t$ is a threshold, $S$ is a step size
of each iteration. $S$ satisfies $0.8 < S < 1$. For the $i^{th}$ iteration, if
$t_i$ satisfies Eq. (6):

\begin{equation}
\frac{Area_0(t_i)}{Area_1(t_1)} < F \& N(t_i) < N, \label{eq6}
\end{equation}

where $t_1$ is the determined threshold value.

Fig. 3 illustrates the effectiveness of LI Otsu’s method
using an example image with the muscle edge. Fig. 3(a) shows
the intensity value histograms of an image which gray scale
image variant is shown in Fig. 3(b). The LI Otsu’s method
obtains the threshold value as 118 producing the segmentation
result in Fig. 3(d). To compare, the Otsu’s method yields the
threshold value 146 that gives the segmentation result shown in
Fig. 3(c). It can be observed that unlike its counterpart, the LI
Otsu’s method retains only WBCs.

\section{MUSCLE EDGE DETECTION}

Muscle edge detection is performed in two substeps:
muscle texture detection and fuzzy WBC detection. Based on
the muscle texture, the muscle edge can be detected. In a
situation when WBCs are in high density, the muscle texture
could be hardly recognized. Fuzzy classification method allows
for refining the result. To accurately detect the muscle edge,
the results generated from muscle texture detection and fuzzy
WBC classification are merged. Fig. 4 illustrates the steps of
the two detectors. The highlighted area shows the necessity of
two detections. Due to the densely distributed WBCs in the red
rectangle, muscle texture detection returns a result as no texture
in this area. This error can be fixed by combining the result of
fuzzy WBC detection.

\subsection{Muscle texture detection}

Consider the muscle image $f(x, y)$ which represents the
intensity function of the input image. The step-by-step
procedure of muscle texture detection is described below.

\textit{1) Apply 16*16 average filter to reduce the appearance of
cells and strengthen the muscle fiber texture.}

\begin{equation}
g(x, y) = f(x, y)\cdot h_{ave}(x, y)
\end{equation}

\textit{2) Utilize Gaussian smoothening filter to smooth the
image and filter out the noise.}

\begin{equation}
g_2(x, y) = g(x, y)\cdot G_{\sigma}(x, y)
\end{equation}

\textit{3) Apply sharpening filter to strengthen texture in muscle
area.}

\begin{equation}
g_3(x, y) = g_2(x, y)\cdot G_{\sigma}(x, y)
\end{equation}

\begin{equation}
g_4(x, y) = k[ g_2(x, y)- g_{3}(x, y)]
\end{equation}

\textit{4) Apply histogram equalization to evenly distribute the
intensity of the image.}

\begin{equation}
p_n = \frac{N_i}{J\cdot K}
\end{equation}

\begin{equation}
g_5(x,y) = floor[(L-1)\Sigma_{n=0}^{f(x,y)}p_n]
\end{equation}

where $N_i$ is the number of pixels with intensity $I$, $i= 0, 1,
…, L-1$; $J, K$ are the width and length of the input image,
$floor()$ rounds down to the nearest integer.

\textit{5) Threshold:}

Standard threshold is applied to $g_4$ output image. However,
the threshold value t is obtained using Eq. (4) over $g_5$. Let $g_6$ be
the result of threshold.

\begin{equation}
g_6 = 
\begin{cases}
  1 & \text{if $g_5(x,y)>t$} \\
  0 & \text{otherwise}
\end{cases}
\end{equation}

\textit{6) Perform image inverse and morphological closing
function.}

\begin{equation}
g_7(x,y) = [(g_6^{-1}\Theta b)\oplus b](x,y),
\end{equation}
where $b(x, y)$ is the structuring element.

\textit{7) Perform corner detection to include edges appearing
within 20x20 pixel area to any corner of the image.} Based on
image capturing assumption, muscle edge must be laid on the
edge or corner of the input image. Therefore, if a candidate
does not contain a pixel within the 20x20 pixel area to any
corner of the image, it is excluded. Also, any object that is
smaller than 60,000 pixels is excluded.

\subsection{Fuzzy WBC detection}

The fuzzy WBC detector does not include average filter
and sharpening filter as the muscle texture detector does.
Therefore, this detector is able to capture WBCs and reveal
void space (the part of image that does not contain any muscle
area). A step-by-step procedure of fuzzy WBC detection is
described below.

\textit{1) Calculate mean of f(x, y) as:}

\begin{equation}
Mean = \frac{1}{N} \Sigma_{x,y}^{J,K}f(x,y),
\end{equation}
where $J, K$ are the width and the length of the image, and $N$
is the total number of pixels.

\textit{2) Apply Gaussian smoothening filter to reduce noise.}

\begin{equation}
g(x,y) = f(x,y) \cdot G_\sigma(x,y)
\end{equation}

\textit{3) Threshold:} Standard threshold is applied to $f(x, y)$
output image. The threshold value is obtained by $k_1 \cdot Mean$,
where $k_1$ is a constant. Let $g_2$ be the result of this step.

\begin{equation}
g_2 = 
\begin{cases}
  1 & \text{if $g(x,y)>k_1 \cdot Mean$} \\
  0 & \text{otherwise}
\end{cases}
\end{equation}

\textit{4) Perform inverse and morphology transform to expand
and connect the effective area}
\begin{equation}
g_3(x,y) = [(g_2^{-1}\Theta b)\oplus b](x,y),
\end{equation}
where $b(x, y)$ is the structuring element.

\textit{5) Corner detection as in muscle texture detection
algorithm.}

\subsection{Intersection}

After the results from two detectors are generated, common
areas are obtained by intersecting two results. Via excluding
small objects and filling holes, an “empty space” is determined.
The size threshold is reduced to 50,000 pixels for fine tuning. A higher threshold value can exclude potential candidates. The
muscle edge can be obtained by applying Canny Edge Detector
[2], as shown in Fig. 4.

\begin{figure}[h] 
    \centering
    \includegraphics[width=0.5\textwidth]{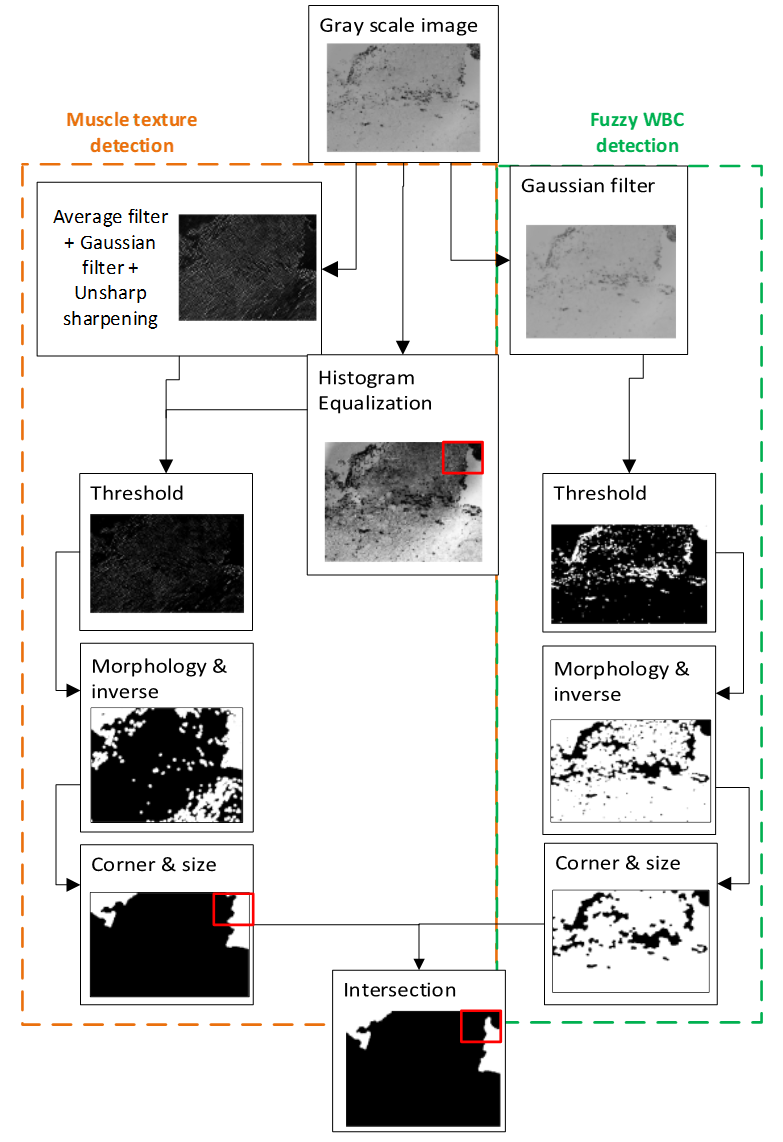} 
    \caption{Muscle edge detection scheme}
    \label{fig:5}
\end{figure}

\begin{figure}[h] 
    \centering
    \includegraphics[width=0.5\textwidth]{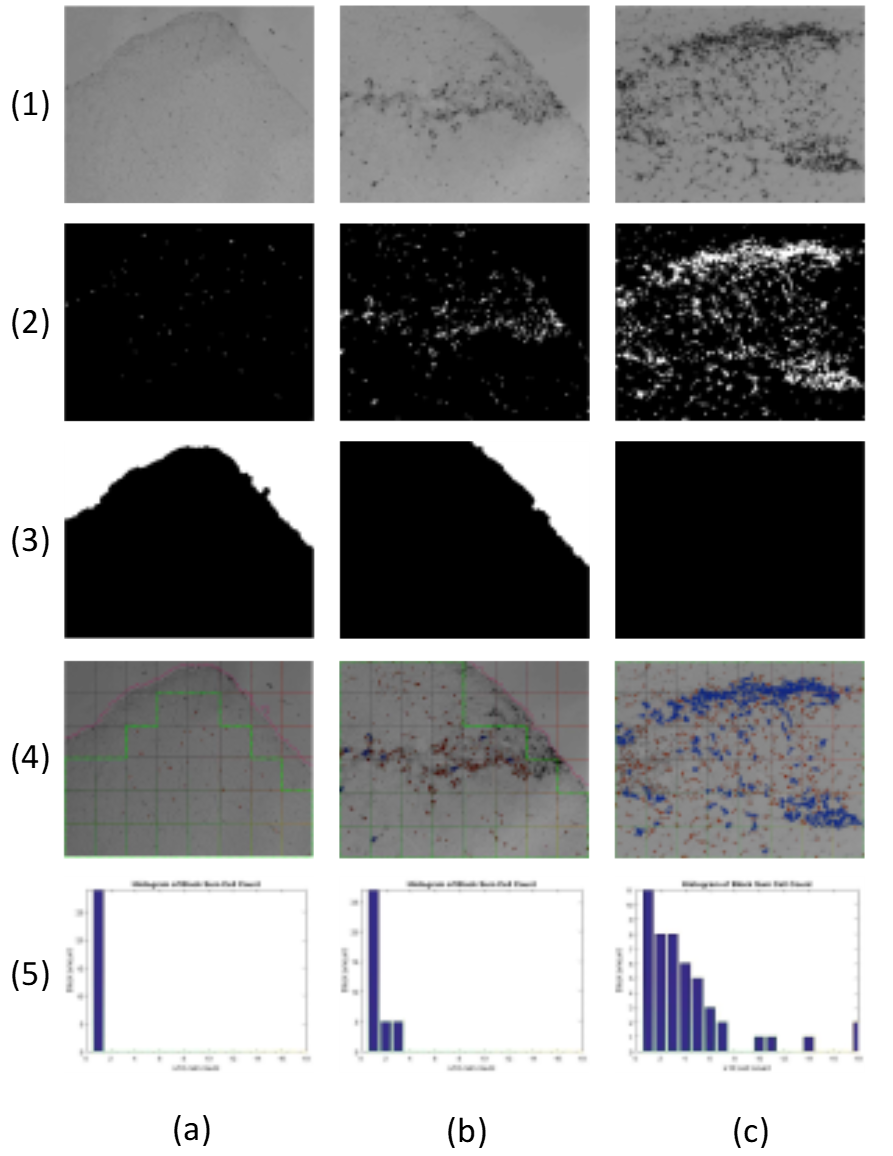} 
    \caption{(1) Gray scale image (2) Threshold result (3) Muscle edge detection (4) Result of segmentation \& ROI (5) Block histogram.}
    \label{fig:6}
\end{figure}

\subsection{ROI selection and block analysis}

After the muscle edge is detected, any object close to the
muscle edge is excluded (with distance d<200 pixel). The
remaining objects are “effective” objects or effective WBCs.
The ROI selection is based on the block unit with the block
size of 200x200 pixels. For each block, we use a score to
determine if the block is part of the ROI. The score is
calculated based on the distance between each corner of a
block and the muscle edge, the number of effective objects and
the percentage of void area in a block.

From the above, objects obtained by Localized Iterative
Otsu’s threshold and located in the ROI are counted.
Background (unwanted) objects outside of the ROI are
excluded. We use examples of three images to demonstrate the
results of edge detection and ROI selection.

In Fig. 5, three groups of image are shown as (a) (b) and
(c). Images of (a) and (b) display a muscle edge. Rows (1) and
(2) demonstrate gray scale images and LI Otsu’s threshold
results. In row (3) we can see that the muscle edge is found by
the proposed muscle edge detection. In row (4), ROIs are
outlined by green dash lines. Red spots show the discrete
WBCs and blue areas show the clusters of WBCs. Row (5)
demonstrates the block histograms. Block histograms depict

the amount of blocks that contain a certain number of WBCs.
The number of WBCs in clusters is estimated based on Eq. (3).
For the image (a) (see, Fig. 5) all 29 effective blocks contain 0-
10 WBCs. For the image (b), 27 blocks contain 0-10 WBCs, 5
blocks have 11-20 WBCs and 5 blocks have 21-30 WBCs. The
image (c) has blocks containing more than 180 WBCs. Thus,
block histograms can demonstrate the progress of the cell
density distribution in time and space.

\begin{figure*}[h] 
    \centering
    \includegraphics[width=1\textwidth]{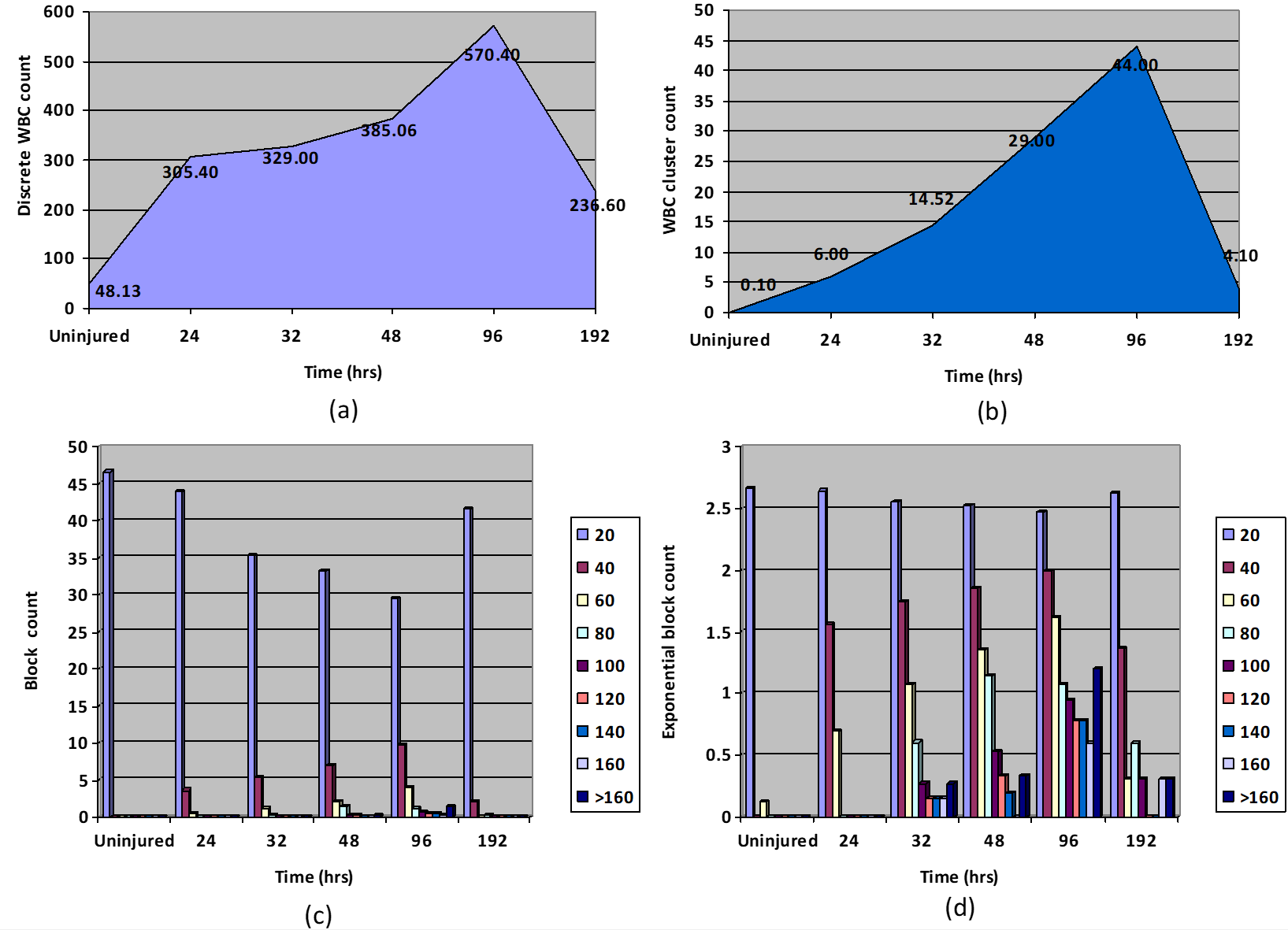} 
    \caption{(a) Discrete white blood cells (b) Clustered white blood cells (c) Block histogram (d) Exponential block histogram}
    \label{fig:7}
\end{figure*}

\begin{table}[h]
\centering
\caption{Performance Comparison}
\scalebox{1.0}{
\begin{tabular}{ |c|c|c|c| } 
 \hline
 \textbf{Total} & \textbf{Max Entropy} & \textbf{Yen} & \textbf{LI Otsu’s} \\ 
 False Positive & 52 & 49 & 42 \\ 
 False Negative & 15 & 24 & 17 \\ 
 Debris & 9 & 6 & 10 \\ 
 Total Counting & 374 & 398 & 370 \\ 
 Empty space resistance & No & No & Yes \\ 
 Accuracy & 79.68\% & 80.15\% & 81.35\% \\ 
 \hline
\end{tabular}
}

\label{table:1}
\end{table}

\begin{table*}[h]
\centering
\caption{Feature Extracted from Muscle Images}
\scalebox{1.0}{
\begin{tabular}{ |c|c|c|c|c|c|c| } 
 \hline
 \textbf{Per image} & \textbf{Uninjured} & \textbf{24-hr} & \textbf{32-hr} & \textbf{48-hr} & \textbf{96-hr} & \textbf{192-hr} \\ 
 \# of Discrete WBCs & 48.13 & 305.40 & 329.00 & 385.06 & 570.40 & 236.60 \\ 
 \# of Clustered WBCs & 0.10 & 6.00 & 14.52 & 29.00 & 44.00 & 4.10 \\ 
 Size of Discrete WBCs (pixels) & 58.90 & 87.10 & 97.96 & 109.63 & 106.98 & 77.21 \\ 
 \# of WBCs in clusters & 2.37 & 68.80 & 177.70 & 341.00 & 1513.00 & 87.20 \\ 
 Overall \# of WBCs & 50.50 & 374.20 & 506.70 & 726.06 & 2083.40 & 323.80 \\ 
 \hline
\end{tabular}
}
\label{table:2}
\end{table*}

\section{Experimental results}

In this section, we analyze WBC expression of the CD68
protein. This analysis utilizes images of mouse cross-sections
of crush-injured gastrocnemius muscle. Injured and uninjured
muscles were collected at 24, 32, 48, 96 and 192 hours after the
injury and then frozen. Each frozen muscle was sliced into 10-
micron thick sections. Serial sections including both injured
and uninjured muscles were stained with an antibody that
recognizes CD68-positive cells. Images were captured under
100X magnification and adjusted by Image-Pro software. Also,
a group of images were obtained from uninjured muscles of
mice that did not undergo crush injury. In the following, the
comparison results of the proposed LI Otsu’s threshold with
the threshold methods in ImageJ are presented before the
quantification results of the proposed framework are discussed.

\subsection{Comparison with other threshold methods}
ImageJ incorporates 16 thresholding methods [18], among
them, Max Entropy and Yen perform the best according to the
results. Therefore, these two methods together with Otsu’s
method are compared with the proposed LI Otsu’s method.
First, uninjured images are processed using these methods. It’s
observed that Otsu’s method produces many false objects. In
Fig. 6, it can be seen that the original image does not contain
any WBC, however the method produces objects. Max Entropy
and Yen methods have 30-50 false positives. The proposed LI
Otsu’s method has only 1 false positive which is the best result.
Secondly, the results from the proposed LI Otsu’s, Max
Entropy and Yen are compared with the ground truth. The
ground truth results include the manual counting results of 12
image blocks. These 12 image blocks are selected from 3 time
points including uninjured case, 24 and 32 hours after injury.
Table 1 shows the quantitative comparison of the result by
each method against the ground truth. The proposed LI Otsu
method has 81.35
accuracy than the Yen method and 1.67
Max Entropy. During the image acquisition process, a low
intensity noise is introduced that can produce false WBCs.
From the table, it can be seen that the proposed LI Otsu’s
method does not cope very well with such kind of noise, which
may be improved further by classification.

\subsection{Quantification result}
The proposed framework is implemented in MATLAB.
Totally, 95 images including 65 images of injured muscles and
30 images of uninjured muscle are processed automatically.
Features presented in Section III are extracted and analyzed,
with the averaged results shown in Table 2.

From Table 2, 48.13 discrete CD68-positive cells and only
0.1 clusters are contained in uninjured muscle. After the injury,
CD68-positive cells and their clusters reach a peak at 96hr and then their amount decrease. After 96hr time point, both discrete
and clustered CD68-positive cellular count decrease. Fig. 7 (a)
shows the trend. A similar trend is observed for discrete WBCs
in Fig. 7 (b). Having a reference from [4], it is confirmed that
the trend of CD68 after injury is true.

The size of discrete CD68-positive cells increases from
uninjured case to 48hr. After 48hr, it presents a decreasing
trend. The number of discrete CD68-positive cells increases
from 48hr to 96hr. We speculate that these WBCs are those
that newly transferred from blood vessels to injured area.
These cells have not been fully activated or in the other word,
have not grown fully mature.

\subsection{Density distribution}

After applying Eq. (3), an estimate of CD68-positive cell
count in clusters and the overall CD68-positive cell count are
obtained (see Table 1). The density distribution of CD68-
positive cells is represented by the binned histogram: the
number of blocks containing $0-20, 21-40, …, >160$ cells are
selected as representative bin values. The histogram of CD68-
positive cell counts per block is shown in Fig. 7. Fig. 7(c)
shows a scaled histogram for the visualization purpose. The
tabulated values using Eq. (19) are presented in Fig. 7(d).

\begin{equation}
N_{block}^{'} = log(10 \cdot N_{block} + 1)
\end{equation}

As shown in Fig. 7 (c), in uninjured muscle, over 99%
blocks contain only 0-20 CD68-positive cells. After the injury,
the number of blocks that contain more than 20 cells increases. At 96 hours post-injury, the density of cells increases in blocks
from 0-20 to >160, that represents the trend of CD68-positive
cell count. Consistent with the trend in Fig. 7 (c) and (d), the
peak density occurs at 96-hr. The density of WBCs decreases
between 96-hr and 192-hr time points, which is reflected by the
decrease of the number of blocks containing more than 20
cells.

\section{Conclusion}

In this paper, an automated quantification and analysis
framework for muscle healing is proposed. The quantified
characteristics include the number and the spatial density of
discrete CD68-positive cells, the number of clustered CD68-
positive cells, and the total number of CD68-positive cells per
selected time point. The developed system automates the
analysis of microscopic images and achieves higher accuracy.
The analysis using the developed procedures show that the
number of CD68-positive cells increases after muscle injury
and reaches the peak at 96 hours after the injury. Future work
would extend the employment of the designed framework to
other protein quantification and the study of correlation among
protein cells during muscle healing. Also, machine learning
method will be introduced to associate with threshold to
achieve more accurate quantification.

\section{Acknowledgement}

This project is supported in part by the UNLV Faculty
Opportunity Award, UNLV Doctoral Graduate Research
Assistantship, and Department of Defense Air Force Grant
(FA-7014-10-2-0001). Review of the material does not imply
Department of the Air Force endorsement of factual accuracy
or opinion.

\bibliographystyle{IEEEtran}
\bibliography{egbib}

\end{document}